\documentclass[amsmath,amssymb,twocolumn]{revtex4}

\usepackage{amsmath}
\usepackage[dvips]{graphicx}
\usepackage{bm}
\usepackage{color}

\begin{document}

\title{Mesoscopic spin Hall effect along a potential step in graphene}

\begin{abstract}
We consider a straight one-dimensional potential step created across a graphene flake. Charge and spin transport through such a potential step are studied in the presence
of both intrinsic and extrinsic (Rashba) spin-orbit coupling (SOC). At normal incidence electrons are completely reflected when the Rashba interaction (with strength $\lambda_R$)
is dominant whereas they are perfectly transmitted if the two types of SOC
are exactly balanced. At normal incidence, the transmission probability of the step is thus controlled continuously from 0 to 1
by tuning the ratio of the two types of SOC. 
Besides the transport of charge in the direction normal to the
barrier, we show the existence of a spin transport 
along the barrier. The magnitude of the spin Hall
current is determined by a subtle interplay between the height of the potential step and the position of Fermi energy. It is demonstrated that contributions from inter-band matrix elements and evanescent modes are dominant in spin transport. Moreover, in the case of vanishing extrinsic SOC ($\lambda_R =0$),
each channel carries a conserved spin current, in contrast to the general case of a finite $\lambda_R$,
in which only integrated spin current is a conserved quantity. Finally, we provide a quasi-classical picture of the charge and spin transport by imaging flow lines over the entire
sample and Veselago lensing (negative refraction) in the case of a $p-n$ junction. 
\end{abstract}

\date{\today}

\pacs{1}

\author{Ai Yamakage$^1$, Ken-Ichiro Imura$^{1,2,3}$, 
J\'er\^ome Cayssol$^{3,4,5}$ and Yoshio Kuramoto$^1$}

\affiliation{$^1$Department of Physics, Tohoku University, Sendai 980-8578, Japan,}
\affiliation{$^2$Department of Quantum Matter, AdSM, Hiroshima University, Higashi-Hiroshima 739-8530, Japan,}
\affiliation{$^3$CPMOH(UMR-5798), CNRS and Universit\'e Bordeaux 1, Talence F-33045, France,}
\affiliation{$^4$Department of Physics, University of California, Berkeley, California 94720, USA,}
\affiliation{$^5$Max-Planck-Institut f\"{u}r Physik Komplexer Systeme, 01187 Dresden, Germany}

\maketitle

\section{Introduction}

Within the last decade the role of SOC in pure crystals has been fundamentaly reconsidered thereby providing a host of 
novel effects \cite{Murakami03,Sinova04,kato04,wunderlich05} and phases \cite{kane05prl1,kane05prl2,BHZ06,HgTe,qi-zhang10physicstoday,moore10,hasan10,qi10prm} with potential applications in spintronics. 

First it was predicted that SOC may generate a transverse 
spin current in response to an applied electric field both in hole and electron doped semiconductors
\cite{Murakami03,Sinova04}. This so-called intrinsic spin Hall effect was promptly reported by the detection of the related spin accumulation at the boundaries of GaAs samples \cite{kato04,wunderlich05}.

More recently SOC was shown to lead to a topological phase of electronic matter when combined with a particular band inversion property \cite{kane05prl1,kane05prl2,BHZ06,HgTe,konig07,roth09,qi-zhang10physicstoday,moore10,hasan10,qi10prm}. In two-dimensional systems this so-called Quantum 
Spin Hall (QSH) state is characterized by metallic edge states surrounding an insulating bulk. Those edge states show up in the absence of magnetic field and realize a time-reversal invariant version of the chiral edge states of the Integer Quantum Hall states. Initially 
the QSH state was introduced in the Kane-Mele model of graphene which describes non-interacting electrons on honeycomb lattice with both intrinsic and extrinsic SOC \cite{kane05prl1,kane05prl2}. In the Kane-Mele model,
intrinsic SOC conserves the $z$-component the real spin ($s_{z}$) and generates a topological mass term. In contrast 
the extrinsic Rashba-type SOC breaks $s_{z}$ conservation by mixing $s_z =+1/2$ (spin $\uparrow$) and $s_z = -1/2$ (spin $\downarrow$) spin components.
When the intrinsic SOC dominates the Rashba one, the bulk excitations are gapped and a pair of gapless states counter-propagate along an edge of the sample. Nevertheless the weakness of the SOC induced gap \cite{min06,huertas06,yao07} makes the realization of the QSH state extremely difficult in graphene.
In contrast similar QSH edge states have been predicted \cite{BHZ06} and soon after reported in transport experiments,
performed in HgTe/CdTe quantum wells \cite{HgTe,konig07,roth09}.

In a recent paper \cite{EPL}, we have shown that charge transport through a potential step allows to investigate the relative strength of intrinsic and extrinsic SOCs.
In the absence of SOC, the transmission of electrostatic potential steps have been
extensively studied in graphene both experimentaly
\cite{huard07, williams07, ozyilmaz07, oostinga07,gorbachev08,liu08} and theoretically \cite{katsnelson06, cheianov06, cayssol09}
with the purpose of providing a condensed-matter implementation of the relativistic Klein tunneling. 
In practise such potential steps can be induced either by a distant gate 
\cite{huard07, williams07, ozyilmaz07, oostinga07, gorbachev08, liu08} 
or by metallic contacts \cite{huard08,huard09}.

In this paper we detail supplementary aspects of the charge transport properties outlined in \cite{EPL}. In addition we emphasize here the spin transport
near an electrostatic potential step. 
As a main result, 
an interfacial spin Hall effect is predicted, 
which we call hereafter mesoscopic spin Hall effect (MSHE).
In the MSHE the spin current flows in the direction transverse to the applied electric field, and is mainly 
localized at the vicinity of the step. Owing to this nonuniform spin current distribution, the MSHE differs from the spin Hall effect in homogeneous doped semiconductors 
\cite{Murakami03, Sinova04, kato04, wunderlich05} and in graphene \cite{dyrdal09}.

The paper is organized as follows. 
In Sec. II, the charge and spin current operators for the Kane-Mele model are derived in presence of intrinsic and
Rashba SO coupling. 
It is shown in section III that quantum averages of these operators consist of direct and cross (interference) terms when
computed within a generic scattering state. 
We discuss charge transport in Section \ref{charge_transport}.
As a main prediction of this paper, the spin transport along
the interface (transverse to the applied electric field) is described
thoroughly in the section V while possible experimental detection is also
discussed therein. Finally, we provide a quasi-classical picture of the charge and spin transport by imaging flow lines on the entire
sample and Veselago lensing (or negative refraction) at the 
$p-n$ junction.

\section{Formalism: charge and spin currents}

We consider a graphene monolayer in the presence of 
both intrinsic and extrinsic SOC . 
The corresponding charge and spin current operators are constructed in the
framework of the Kane-Mele model of graphene \cite{kane05prl1,kane05prl2}. The quantum averages
of the charge and spin currents are derived both for propagative and for
evanescent quasiparticles.

\subsection{Kane-Mele model}

The low-energy Kane-Mele model \cite{kane05prl1,kane05prl2} is defined by the Hamiltonian $%
H_{KM}=H_{0}+H_{SO}+H_{R}$ with 
\begin{eqnarray}
H_{0} &=& \psi^\dag (p_{x}\sigma _{x}\tau _{z}+p_{y}\sigma _{y}) \psi,  \notag \\
H_{SO} &=&-\psi^\dag \Delta \sigma _{z}\tau _{z}s_{z} \psi,  \notag \\
H_{R} &=& \psi^\dag \lambda _{R}(\sigma _{y}s_{x}-\sigma _{x}\tau _{z}s_{y}) \psi.
\label{HKM}
\end{eqnarray}%
Here $\sigma _{i},\tau _{i},s_{i}$ $(i=x,y,z)$ are the Pauli matrices
associated with the lattice isospin ($A$ and $B$ sites of the honeycomb
lattice), the valley-isospin ($K$ and $K^{\prime }$ points of the reciprocal
space), and the real electronic spin, respectively.

The kinetic Hamiltonian $H_{0}$ describes graphene in the absence of any
spin-orbit interaction. The intrinsic spin-orbit effect (with coupling
constant $\Delta $) is described by $H_{SO}$ which preserves\ all the
symmetries of the problem and further conserves the component $s_{z}$ of the
electronic spin. In contrast the Rashba contribution $H_{R}$ (with coupling
constant $\lambda _{R}$) explicitely breaks the conservation of $s_{z}$.

The full Kane-Mele Hamiltonian $H_{KM}$ acts on $8$-spinors. Nevertheless in
this paper we will only consider intravalley scattering caused by
electrostatic potential steps. Hence we shall focus on the $K$-valley in the
following analysis and simply substitute ($\tau _{z}=1$) in Eq.(\ref{HKM}).
The resulting Hamiltonian $H_{KM}^{(K)}$ consists of a $4 \times 4$ matrix acting on
a spinor of the form $^{t}[\psi _{A\uparrow },\psi _{B\uparrow },\psi
_{A\downarrow },\psi _{B\downarrow }]$, where $\uparrow $, $\downarrow $
stands for real spin.

In the homogeneous case (in the absence of spacially varying potential), the
momentum $\bm{p}=(p_{x},p_{y})$ is a good quantum number. The
single-valley Hamiltonian $H_{KM}^{(K)}$ is diagonalized by the eigenstates $%
|\alpha \beta \rangle _{\bm{p}}$ as 
\begin{equation}
H_{KM}^{(K)}|\alpha \beta \rangle _{\bm{p}}=E_{\alpha \beta }(\bm{p}%
)|\alpha \beta \rangle _{\bm{p}},
\end{equation}%
where indices $\alpha ,\beta =\pm 1$ specify the band. The dispersion
relation of the band $\alpha \beta $ reads 
\begin{equation}
E_{\alpha \beta }(\bm{p})=\alpha \sqrt{\bm{p}^{2}+(\Delta +\beta
\lambda _{R})^{2}}+\beta \lambda _{R},  \label{Eab}
\end{equation}%
and the corresponding wavefunction $\psi _{\alpha \beta \bm{p}}(\bm{x%
})=\left\langle \bm{x}|\alpha \beta \right\rangle _{\bm{p}}$ can be
expressed as 
\begin{equation}
\psi _{\alpha \beta \bm{p}}(\bm{x})=A_{\alpha \beta }(\bm{p})%
\left[ 
\begin{array}{c}
p_{x}-ip_{y} \\ 
E_{\alpha \beta }(\bm{p})+\Delta \\ 
-i\beta \left( E_{\alpha \beta }(\bm{p})+\Delta \right) \\ 
-i\beta (p_{x}+ip_{y})%
\end{array}%
\right] e^{i \bm p \cdot \bm x},  \label{Eigen}
\end{equation}%
with 
\begin{equation}
1/A_{\alpha \beta }(\bm{p})
	=\sqrt{2 \left[ 
	\bm{p}^{2}
	+\left( E_{\alpha\beta }(\bm{p})+\Delta \right) ^{2}
	\right]WL}.  \label{norm}
\end{equation}%
These wavefunctions are normalized to represent a unit probability within a
rectangular graphene flake of length $L$ (along the $x$-direction) and width 
$W$ (along the $y$-direction).

In the presence of a potential step or a sample edge, evanescent states appears.
Their spectrum and wave functions are given, respectively, by Eqs. (\ref{Eab},\ref{Eigen},\ref{norm}) with typically an
imaginary $p_{x}$. 
Note also that 
$\bm{p}^{2}=p_{y}^{2}-\left\vert p_{x}\right\vert^{2}$. 
Boundedness of the wave function does not allow
evanescent modes to exist in an ideally infinite system. 
They are, nevertheless, ubiquitous in any hetero-junctions (near potential steps) or even at a sample termination. 
The helical edge states, on the other hand, occur
only at the boundary between two topologically 
distinguishable phases each characterized by opposing $Z_2$ indices. In the case of $Z_{2}$ topological insulator,
$Z_2$ index distinguishes trivial (even number of
Kramer's pairs) vs. non-trivial (odd number of Kramer pairs) insulating states.

\subsection{Charge and spin current operators}

We construct here the charge and spin current operators associated with the
single-valley Kane-Mele Hamiltonian $H_{KM}^{(K)}$. We simply write the
continuity equations for the charge and the $s_{z}$-component of the
electronic spin. The charge and spin density operators ($\rho _{\mathrm{c}}(%
\boldsymbol{x})$ and $\rho_{s_z}(\boldsymbol{x})$ respectively) can be
written as, 
\begin{align}
\rho _{\mathrm{c}}(\boldsymbol{x})& =-e\psi ^{\dag }(\boldsymbol{x})\psi (%
\boldsymbol{x}), \\
\rho _{s_{z}}(\boldsymbol{x})& ={\frac{\hbar }{2}}\psi ^{\dag }(\boldsymbol{x%
})s_{z}\psi (\boldsymbol{x}),
\end{align}%
using the four component field operators $\psi (\boldsymbol{x})$ and $\psi
^{\dag }(\boldsymbol{x})$. The charge current operator $\boldsymbol{J}_{%
\mathrm{c}}$ is determined such that it satisfies the continuity equation
for charge: 
\begin{equation}
{\frac{\partial \rho _{\mathrm{c}}}{\partial t}}+\mathrm{div}\boldsymbol{J}_{%
\mathrm{c}}=0.  \label{div_c}
\end{equation}%
Using the equation of motion 
\begin{equation}
i{\frac{\partial \psi }{\partial t}} =\left[ -i\boldsymbol{\sigma }\cdot 
\boldsymbol{\nabla } +\lambda _{R} \left( \sigma_y s_x - \sigma_x s_y
\right) -\Delta \sigma _{z}s_{z}) \right]\psi ,  \label{EOM}
\end{equation}%
it is readily verified that the operator defined as, 
\begin{equation*}
\boldsymbol{J}_{c}\equiv -e\psi ^{\dag }\boldsymbol{\sigma }\psi ,
\end{equation*}%
satisfies Eq. (\ref{div_c}). This expression can be alternatively obtained
from the usual definition of the charge current, 
\begin{equation*}
\boldsymbol{J}_{c}= -e\frac{\partial H_{KM}^{(K)}}{\partial \bm{p}},
\end{equation*}%
and by noting that the only momentum-dependent part of $H_{KM}^{(K)}$ is $%
H_{0} = \psi^\dag \bm{p}  \cdot \bm \sigma \psi$, ($v_{F}=1$).

We repeat the same procedure by introducing the spin current operator for $%
s_{z}$-spin component as 
\begin{equation}
\boldsymbol{J}_{s_{z}}={\frac{\hbar }{2}}\psi ^{\dag }\boldsymbol{%
\sigma }s_{z}\psi .  \label{js_def}
\end{equation}%
However, in the presence of Rashba SO coupling, the second term on the
r.h.s. of Eq. (\ref{EOM}) does not commute with $s_{z}$, leading to a source
term in the continuity equation for spin: 
\begin{equation}
{\frac{\partial \rho_{s_{z}}}{\partial t}}+\mathrm{div}\boldsymbol{J}_{%
{s}_{z}}=\lambda _{R} \psi^\dag \left(\sigma _{x}s_{x}+\sigma _{y}s_{y}\right)\psi .
\label{continuity_s}
\end{equation}%
The source term $\lambda _{\mathrm{R}}(\sigma _{x}s_{x}+\sigma _{y}s_{y})$
describes the spin torque due to Rashba spin orbit interaction. Violation of
the continuity equation leads in many cases to the ambiguity in the
definition of spin current density, reflecting correctly the fact that spin
is not conserved. However, as far as spin transport in our junction problem
is concerned, we will see later in Sec. \ref{spintransport} that the spin current density
defined as Eq. (\ref{js_def}) yields a conserved spin current when
integrated over the incident angle $\phi $. Such a spin current satisfies
the continuity equation \textit{globally}, i.e. in the sense of%
\begin{equation}
\int_{-p_{F}}^{p_{F}}\frac{dp_{y}}{2\pi }\left( {\frac{\partial \left\langle \rho _{s_{z}}%
\right\rangle}{\partial t}}+\mathrm{div} \left\langle \bm{J}_{s_{z}} \right\rangle \right) =0,
\label{spinconservation}
\end{equation}%
where $\left\langle ... \right\rangle$ denotes averaging over the scattering states. The spin current density (\ref{js_def}) becomes \textit{locally} conserved
in the absence of Rashba SO coupling: $\lambda _{R}=0$.

\subsection{Charge and spin current carried by an eigenstate $|\protect%
\alpha \protect\beta \rangle _{\bm{p}}$}

We now evaluate the quantum average of the charge and $s_{z}$-spin currents
in the eigenstate $|\alpha \beta \rangle _{\bm{p}}$ defined by Eq. (\ref%
{Eigen}). When both $p_{x}$ and $p_{y}$ are real, this state corresponds to
a propagating plane wave and carries the charge current density 
\begin{align}
\left\langle \bm{J}_{\mathrm{c}}\right\rangle _{\alpha \beta \bm{p}%
}& =-e \psi _{\alpha \beta \bm{p}}^{\dagger }(\bm{x})\bm{\sigma 
}\psi _{\alpha \beta \bm{p}}(\bm{x})  \notag \\
& =-4eA_{\alpha \beta }^{2}(E_{\alpha \beta }(\bm{p})+\Delta )\ \bm{p%
}  \label{jc_bulk}
\end{align}%
which is collinear to the momentum $\bm{p}$. This current consists in
equal contributions from the channels $s_{z}=1$ and $s_{z}=-1$. Moreover
these contributions correspond to opposite spin currents leading to a
cancellation of spin current for the $s_{z}$-spin component. This statement
is confirmed by direct calculation of the quantum average of the $s_{z}$%
-spin current $\bm{J}_{\mathrm{s}_{z}},$%
\begin{equation}
\left\langle \bm{J}_{\mathrm{s}_{z}}\right\rangle _{\alpha \beta \bm{%
p}}
	=\psi _{\alpha \beta \bm{p}}^{\dagger }(\bm{x})\bm{\sigma
}s_{z}\psi _{\alpha \beta \bm{p}}(\bm{x})
	=\bm{0,}
\label{js_bulk}
\end{equation}%
in the bulk eigenstate $|\alpha \beta \rangle _{\bm{p}}$.

In presence of a potential step or at a sample edge, evanescent states,
described by an imaginary $p_{x}$, become possible. If we assume a
semi-infinite graphene plane extending over the half-plane ($x>0$), the
charge current carried by such an evanescent wave reads%
\begin{align}
\left\langle J_{\mathrm{c}}^x\right\rangle _{\alpha \beta \bm{p}}& =0,
\label{jc_bulkeva} \\
\left\langle J_{\mathrm{c}}^y\right\rangle _{\alpha \beta \bm{p}}&
=-4eA_{\alpha \beta }^{2}(E_{\alpha \beta }(\bm{p})+\Delta )\
p_{y}e^{-2\left\vert p_{x}\right\vert x}.
\label{jc_bulkevay}
\end{align}%
The current is localized near the interface and flows along the $y-$axis.
Moreover the net charge transport vanishes when the sum over $p_{y}$ is
performed. 

Finally the average of the spin current in an evanescent state, 
\begin{align}
\left\langle J_{\mathrm{s}_{z}}^x\right\rangle _{\alpha \beta 
\bm{p}}& =0,  \label{js_bulkeva} \\
\left\langle J_{\mathrm{s}_{z}}^y\right\rangle _{\alpha \beta 
\bm{p}}& =-2A_{\alpha \beta }^{2}(E_{\alpha \beta }(\bm{p})+\Delta
)\ \left\vert p_{x}\right\vert e^{-2\left\vert p_{x}\right\vert x},
\label{jszy_direct}
\end{align}%
shares the characteristics of the charge current except that spin current will
not vanish upon $p_{y}$ integration (see section V).

\section{Scattering states for a potential step}

\begin{figure}[tbp]
\includegraphics[width=8cm]{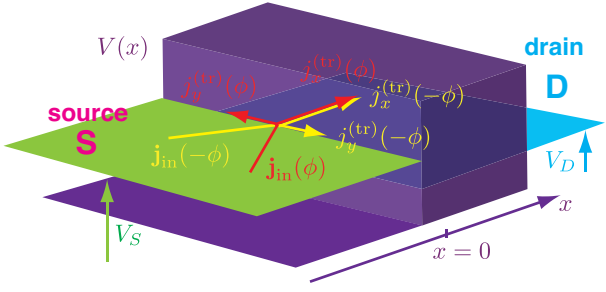}
\caption{Potential step $V(x)$.}
\label{Fig1setup}
\end{figure}

We consider a potential step in a graphene monolayer as shown in Fig. \ref{Fig1setup} and construct the
corresponding scattering states. We evaluate the quantum averages of the
charge and spin current densities in a given scattering state. Owing to the
spin-orbit coupling (i.e. multiband character of the Kane-Mele model), the
structure of those averages is more complicated than the average currents in
a pure eigenstate $|\alpha \beta \rangle _{\bm{p}}$. Indeed a single
incident electron generates two transmitted electronic waves. Therefore the
averaged currents contain direct terms involving only one kind of quasiparticles, and crossed terms describing coherent interferences 
between the two transmitted waves.

\subsection{The junction model}

We consider an electrostatic gate creating a potential step $V(x)$ which
is slowly varying on the scale of the atomic lattice. Moreover we assume an ideally pure system which can be approached in experiments with suspended
devices. These conditions ensure that intervalley scattering can be
neglected and that the junction is described by the Hamiltonian 
\begin{equation}
H=H_{KM}^{(K)}+ \psi^\dag V(x)\sigma _{0} s_{0} \psi,
\end{equation}%
where $\sigma _{0}$ and $s_{0}$ represent the identity in lattice isospin
and real spin space, respectively.\ Besides, we further assume that the step
is\ sharp on the scale of the Fermi wavelength. In this sense we may
represent the potential $V(x)$ by an abrupt step defined by 
\begin{equation}
V(x)=\left\{ 
\begin{array}{ll}
0 & (x<0) \\ 
V_{0} & (x>0)%
\end{array}%
\right. .  \label{step}
\end{equation}%
Note that we also assume a straight interface with translational invariance
along the $y$ direction (no roughness along the interface $x=0$).

\subsection{Scattering states}

Typical scattering state with energy $E$ and transverse momentum $p_{y}$ 
\begin{equation}
\Psi _{E,p_{y}}(\bm{x})=\left\{ 
\begin{array}{ll}
\psi _{I}(\bm{x})+r\psi _{R}(\bm{x})+r_{ev}\psi _{ev}(\bm{x}) & 
(x<0) \\ 
t_{+}\psi _{+}(\bm{x})+t_{-}\psi _{-}(\bm{x})\equiv \psi _{T}(%
\bm{x}) & (x>0)%
\end{array}%
\right. ,  \label{Psi_LR}
\end{equation}%
can be constructed in terms of the bulk eigenstates Eq. (\ref{Eigen}). It is
assumed here that the incident electron is a pure $|+-\rangle _{\bm{p}}$
state. Hence the incident and reflected waves read 
\begin{equation*}
\psi _{I,R}(\bm{x})=A(E,\bm{p})\left[ 
\begin{array}{c}
\pm p_{x}-ip_{y} \\ 
E+\Delta  \\ 
i\left( E+\Delta \right)  \\ 
i(\pm p_{x}+ip_{y})%
\end{array}%
\right] e^{i(\pm p_{x}x+p_{y}y)},
\end{equation*}%
where $p_{x}$(resp. $-p_{x}$) corresponds to the incident $\psi _{I}$ (resp.
reflected $\psi _{R}$) wave, and 
\begin{equation*}
1/A(\bm{p})
	=\sqrt{2 \left[\bm{p}^{2}+\left( E+\Delta \right) ^{2}
	\right]WL}.
\end{equation*}%
The $x$-component of the momentum $p_{x}=p_{x}(E,p_{y})$ is the positive
solution of \ $E_{+-}(p_{x},p_{y})=E$. There is also an evanescent state defined in the half-plane $x<0$, 
\begin{equation*}
\psi _{ev}(\bm{x})=A(E,\bm{p})\left[ 
\begin{array}{c}
-i\kappa -ip_{y} \\ 
E+\Delta  \\ 
i\left( E+\Delta \right)  \\ 
i(-i\kappa +ip_{y})%
\end{array}%
\right] e^{\kappa x+ip_{y}y},
\end{equation*}%
and localized near the interface $x=0$. The value of $\kappa =\kappa (E,p_{y})$ is set 
by the positive solution of $E_{+-}(i\kappa ,p_{y})=E$

Within the half-plane $x>0$, the scattering state consists of two
transmitted waves with opposite symmetry $\beta =+$ and $\beta =-$,
described by the spinors%
\begin{equation*}
\psi _{\beta }(\bm{x})=A(E,\bm{p})\left[ 
\begin{array}{c}
p_{\beta x}-ip_{y} \\ 
E+\Delta \\ 
-i\beta \left( E+\Delta \right) \\ 
-i\beta (p_{\beta x}+ip_{y})%
\end{array}%
\right] e^{i(p_{\beta x}x+p_{y}y)},
\end{equation*}%
where the $x$-components of the momentum $p_{\beta x}=p_{\beta x}(E,p_{y})$
are obtained by solving $E_{\alpha \beta }(p_{\beta x},p_{y})=E-V_{0}$. Note
that $p_{\beta x}$ can be either real or imaginary. When $p_{\beta x}$ is
real, $\psi _{\beta }$ stands for a propagating mode and the sign of $p_{x}$
is chosen such that the group velocity is positive, i.e. correctly describes
an outgoing wave. When $p_{\beta x}$ is imaginary, $\psi _{\beta }$ is an
evanescent mode and the sign of the imaginary part of $p_{\beta x}$ is fixed
by the requirement of wavefunction boundness at $x\rightarrow \infty $.

It is sometimes more convenient to specify $\Psi _{E,p_{y}}(\bm{x})$ by $%
E$ and an incident angle $\phi $, instead of $E$ and the transverse momentum 
$p_{y}:$ 
\begin{equation}
p_{x}=\left\vert \bm{p}\right\vert \cos \phi ,\text{ \ }p_{y}=\left\vert 
\bm{p}\right\vert \sin \phi .
\end{equation}

Finally the four scattering amplitudes $r$, $r_{ev}$, $t_{+}$ and $t_{-}$
are uniquely determined by solving the continuity condition at $x=0$ \cite{EPL}: 
\begin{equation}
\Psi _{E,p_{y}}(x=0^{-},y)=\Psi _{E,p_{y}}(x=0^{+},y)  \label{cc}
\end{equation}%
for given $E$, $V_{0}$ and $p_{y}$(or $\phi $).

\subsection{Direct and crossed terms}

We have seen that the scattering state $\Psi _{E,p_{y}}(\bm{x})$ takes
the form of a superposition of two bulk states with opposite band symmetry $%
\beta $. The expectation value of current density in such scattering state 
\begin{align}
\bm{J}_{\mathrm c}& 
	=-e \psi _{T}^{\dagger }\bm{\sigma }\psi _{T}  \notag \\
& =-e \left(|t_{+}|^{2}\psi _{+}^{\dagger }\bm{\sigma }\psi
_{+}+|t_{-}|^{2}\psi _{-}^{\dagger }\bm{\sigma }\psi _{-}   \right)
	\notag \\
& \quad +2\Re \lbrack t_{+}^{\ast }t_{-}\psi _{+}^{\dagger }\bm{\sigma }\psi
_{-}]),  \label{d-c}
\end{align}%
has naturally two types of contributions: direct and crossed terms. The direct
terms (proportional to $|t_{+}|^{2}$ and $|t_{-}|^{2}$) are similar to the
ones addressed in the previous section, see Eqs. (\ref{jc_bulk},\ref{js_bulk},\ref{jc_bulkeva},\ref{jc_bulkevay},\ref{js_bulkeva},\ref{jszy_direct}). In particular such direct terms were shown to carry no
net $s_{z}$-spin current when associated with a propagative wave (Eq.(\ref%
{js_bulk})) whereas evanescent waves carry a finite spin current (Eq.(\ref%
{js_bulkeva})). In contrast the crossed terms $2\Re \lbrack t_{+}^{\ast
}t_{-}\psi _{+}^{\dagger }\bm{\sigma }\psi _{-}]$ always contribute to
spin transport regardless of the nature of the two interfering transmitted waves.

In the following we focus on the charge and spin transport associated with
these latter crossed terms which mix the transmitted waves $\psi _{+}$ and $%
\psi _{-}$ altogether. The crossed charge current is proportional to the
expression 
\begin{align}
\label{cross_c}
&-i\psi _{+}^{\dagger }\bm{\sigma } \psi _{-}
 \nonumber \\
& \quad =2\tilde{A}_{+}\tilde{A}_{-}(\tilde{E}+\Delta )
\begin{bmatrix}
0 \\ 
p_{-x}-p_{+x}^{\ast }%
\end{bmatrix}
 e^{i(p_{-x}-p_{+x}^{\ast })x},
\end{align}%
where $\tilde{E}=E-V_{0},$ while $p_{+x}$ and $p_{-x}$ were defined in the previous
subsection. These crossed terms yield a charge current along the $y$-axis.
In contrast to the direct terms, the crossed current has a spacial
dependence upon the coordinate $x$ (dependence upon $y$ is forbidden by
translational invariance). The total current is therefore divergenceless.

We now consider the spin current. The crossed terms are proportional to: 
\begin{align}
& \psi _{+}^{\dagger }\bm{\sigma } s_{z}\psi _{-}  \notag \\
& =2\tilde{A}_{+}\tilde{A}_{-}(\tilde{E}+\Delta )\left[ 
\begin{array}{c}
p_{-x}+p_{+x}^{\ast } \\ 
2p_{y}%
\end{array}%
\right] e^{i(p_{-x}-p_{+x}^{\ast })x}.  \label{cross_s}
\end{align}%
The spatial distribution of the spin current is oscillatory if both
transmitted states ($\psi _{+}$ \textit{and} $\psi _{-}$) are propagative.
If one of the transmitted states is propagative while the other is
evanescent, the spin current distribution shows damped oscillations.

The $x$-component of Eq. (\ref{cross_s}) is generally finite and has a $x$%
-dependence. The $y$-component of Eq. (\ref{cross_s}) is also finite, but
does not contribution to the divergence of spin current density. Therefore, 
\begin{equation}
\mathrm{div}\bm{J}_{s_{z}}={\frac{\partial J_{s_{z}}^{x}}{\partial x}}%
\neq 0,  \label{div_sne0}
\end{equation}%
remains finite due to the cross term. Recall that a spin current-density is
generally not a conserved quantity, see Eq. (\ref{continuity_s}). In
contrast the contribution from direct terms to the spin current is \textit{%
divergenceless}. A similar circumstance also occurs in a more conventional
semiconductor-based spin Hall system \cite{SM}.

\section{Charge transport}
\label{charge_transport}

We consider the charge transport across an electrostatic potential step, in
the presence of spin-orbit coupling. The charge conductance is directly
determined by the transmission probability whose energy dependences are
investigated thoroughly in this section.

\subsection{Pseudo-reflection symmetry}
The continuum limit of the Kane-Mele model, defined as in Eq. (\ref{HKM}) has a pseudo-reflection symmetry (PRS) operating within each valley.
PRS with respect to the $x$-axis: $y \to -y$ is expressed as
$U = \sigma_x s_y \mathcal P$, where $\mathcal P$
represents a parity operator in two spatial dimensions,
i.e.,$\bm p \rightarrow \bm p' 
= \mathcal P \bm p \mathcal P^{-1} = (p_x,-p_y)$.
The eigenstate of the Kane-Mele Hamiltonian, Eq. (\ref{HKM}),
is also an eigenstate of PRS operator $U$, since
$[H,U]=0$, with an eigenvalue, $-\beta$, i.e., 
$U\psi_{\alpha\beta\bm p}(\bm x) = 
- \beta \psi_{\alpha\beta \bm p'}(\bm x')$.
This can be checked explicitly using Eq. (\ref{Eigen}).
We demonstrate that
PRS is a convenient tool for clarifying
the dependence of reflection and transmission coefficients 
on incident angles.

\subsection{Transmission at normal incidence}

The continuity equation at $x=0$ reads,
\begin{align}
\psi_I + r\psi_R + r_{ev}\psi_{ev} = t_+ \psi_+ + t_- \psi_-,
\label{+py}
\end{align}
where $y$-component of momentum is $+p_y$,
determining the incident angle $\phi$ for a given $p_x$.
Applying PRS operator $U$ from the left to
the above equation, one finds, 
\begin{align}
 \psi_I' + r \psi_R' - r_{ev} \psi_{ev}' = -t_+ \psi_+' + t_- \psi_-'.
\label{-py}
\end{align}
This can be regarded as the continuity equation for momentum$-p_y$, where $\psi' = \psi|_{p_y = -p_y}$.
Comparing the two equations, Eqs. (\ref{+py}) and (\ref{-py}),
one can convince oneself that 
reflection and transmission coefficient are either an even or an odd function of the incident angle:
\begin{eqnarray}
&& r' = r, \ \ \
 r_{ev}'=-r_{ev},
 \nonumber \\ 
&& t_+' = -t_+,\ \ \ 
 t_-' = t_-.
\end{eqnarray}
At normal incidence: $p_y=0$,
two states with an opposing PRS eigenvalue, $-\beta$ are orthogonal to each other at the \textit{spinor level}, i.e.,
\begin{equation}
\psi^\dag_{\alpha\beta (p_x,0)}(\bm 0) \psi_{\alpha'\beta' (p_x',0)}(\bm 0) \propto \delta_{\beta\beta'},
\end{equation}
for arbitrary $p_x$ and $p_x'$.
Consequently, the continuity condition Eq. (\ref{cc}) is decoupled to two equations:
\begin{align}
\label{rccp}
r_{ev} \psi_{ev}	&= t_+ \psi_+, \\
\psi_I + r \psi_R	&= t_- \psi_-,
\label{rccm}
\end{align}
as far as $p_y=0$ at the interface $x=0$.
The reflection coefficient $r$ is determined only by the transmitted state with $\beta = -$, therefore the normal incident charge transport is independent of states of the symmetry different from the incident state. Namely, the transition from $\beta = -$ to $\beta = +$ is impossible due to different symmetry.

Solving Eqs. (\ref{rccp}) and (\ref{rccm}) yields the reflection coefficient
\begin{align}
r = \frac{X-Y}{X+Y},
\end{align}
where $X = p_F(E - V_0 + \Delta)$ with $p_F$ being the Fermi momentum in the incident side, $Y = p_{-x} (E+\Delta)$.
Reflection probability is given by $R = 1-|r|^2$ \cite{EPL}.
Imaginary $p_{-x}$ which corresponds to the evanescent state $|--\rangle$ leads to perfect reflection $R=1$ even if the other transmitted state $|-+\rangle$ is propagating.
This perfect reflection occurs provided $\Delta - 2 \lambda_R < E-V_0 < -\Delta$ in the dominant Rashba  case ($\lambda_R > \Delta$).
By contrast, the perfect transmission $r=0$ occurs
at the phase boundary $\lambda_R = \Delta$.
Therefore the crossover from perfect reflection to perfect transmission occurs with tuning Rashba SO coupling.

\begin{figure}
\begin{center}
\includegraphics{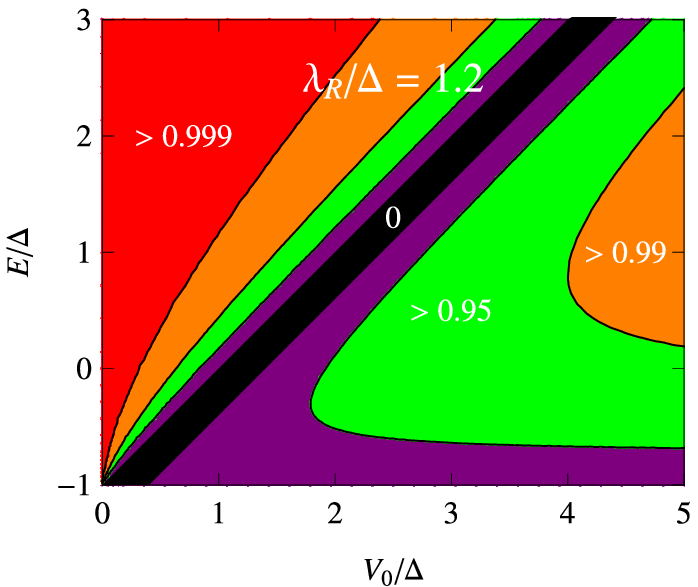}
\end{center}
\begin{center}
\includegraphics{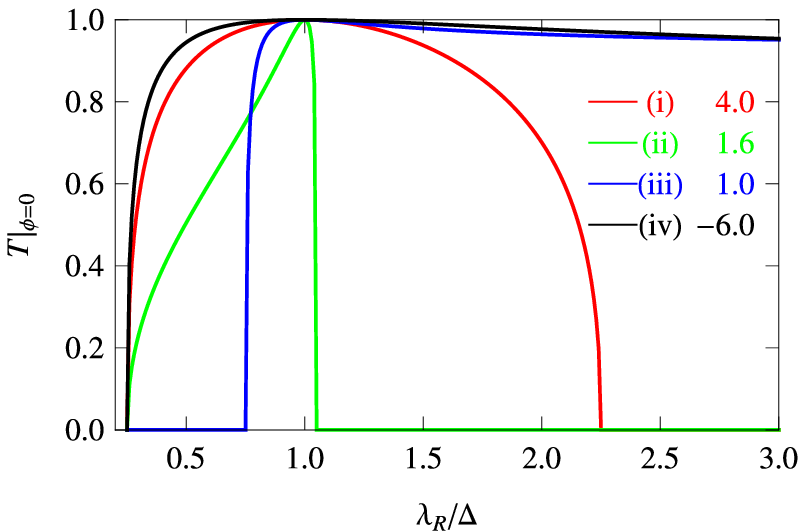}
\end{center}
\caption{Transmission probability $T(p_y=0)$ at the normal incidence
as a function of $V_0/\Delta$ and $E/\Delta$ at $\lambda_R/\Delta = 1.2$
(upper), and 
the dependence of $T(p_y=0)$ on $\lambda_R/\Delta$ 
for a fixed value of $(E/\Delta, V_0/\Delta)=(0.5, 2)$ (lower).
In the upper panel,
each region corresponds to a finite range of $T(p_y=0)$ 
indicated on the panel.
In the lower panel,
each curve corresponds to different values of $V_0/\Delta$.
}
\label{rpcolor}
\end{figure}

The upper panel of Fig. \ref{rpcolor} shows the normal incident transmission probability $T$ as a function of $V_0$ and $E$. The (black) diagonal strip is the region of perfect reflection, which occurs due to dominant Rashba effect. With fixing $V_0, E$ and decreasing $\lambda_R/\Delta$, the reflection probability decreases as shown in the lower panel of Fig. \ref{rpcolor}. In the figure all the lines go through the point being $\lambda_R/\Delta = 1, T|_{\phi=0}=1$, namely perfect transmission \textit{always} occurs, because the Dirac cone appears again at the balanced Rashba and intrinsic SO case ($\lambda_R/\Delta = 1$). 
Further decreasing $\lambda_R/\Delta$, the transmission probability decreases and reaches to zero at $\lambda_R/\Delta = (1-E/\Delta)/2 = 0.25$. Crossover from perfect reflection to perfect transmission occurs due to the competition between Rashba and intrinsic SO in the normal incident case. 

\subsection{Charge conductance}
\label{charge_conductance}

\begin{figure}
\includegraphics{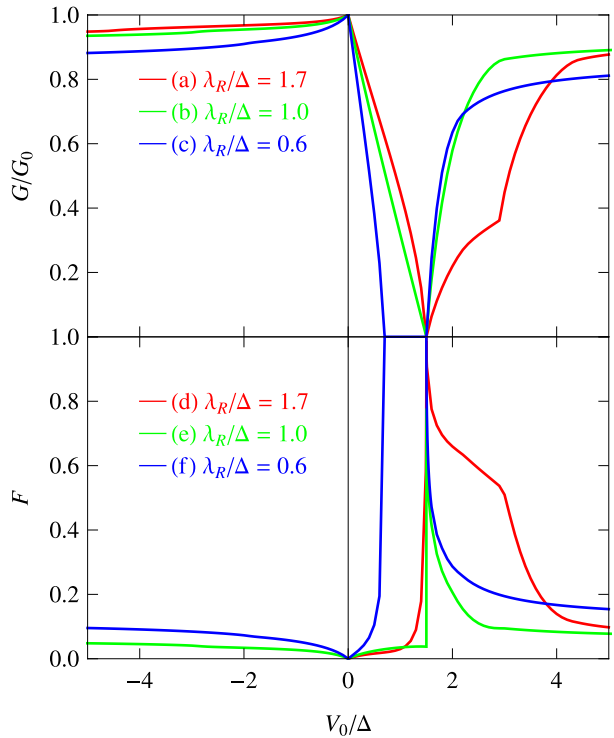}
\caption{Upper panel: charge conductance $G_{c}$ of the potential step (height $V_0$) normalized by $ G_{0}=Wp_F/\protect\pi $ ($E/\Delta =0.5$).
Lower panel: Fano factor $F$.
Red, blue and green curves represent a typical behavior of charge
conductance, respectively, in the semi-metallic phase ($\protect\lambda %
_{R}/\Delta=1.7$), in the topological gap phase ($\protect\lambda _{R} / \Delta = 0.6$) and at
the phase boundary ($\protect\lambda _{R}/\Delta=1$).}
\label{gc}
\end{figure}

Let us consider the rectangular geometry (Fig. \ref{Fig1setup}) with infinite aspect ratio $W/L$, $W$ and $L$  
being respectively the width  and length. We have considered so far the charge current
carried by a single scattering channel with definite $p_{y}$. In the absence of disorder, the channels are independent and 
the charge conductance $G_{c}$ (in units of $e^2/h$) is simply the sum of the single-channel transmissions $T(p_y)$ over all possible transverse momenta $p_{y}$, or equivalently over all incidence angles:
\begin{align}
G_{c}
& =\frac{e^{2}}{h}\int_{-Wp_F}^{Wp_F}\frac{d(Wp_{y})}{2\pi }
T(p_y).
\label{LF}
\end{align}%
We evaluate Eq. (\ref{LF}) explicitly by substituting 
$T(p_y)=1-|r(p_y)|^{2}$, where the reflection amplitude $r(p_y)$ follows from the continuity condition (\ref{cc}). 
Note $T \ne |t_+|^2 + |t_-|^2$ because of potential difference $V_0$ between the incident and transmitted side.
The obtained charge conductance is shown in Fig. \ref{gc} as a function of $V_0$
for different values of $\lambda _{R}/\Delta $. The curves exhibit
specific features depending on the value of $\lambda _{R}/\Delta $. We leave
further inspection of such behaviors to Sec. III C.

Fig. \ref{gc} shows several conductance curves as a function of $V_0/\Delta $
for different values of $\lambda _{R}$. The incident energy is set to be $%
E/\Delta =0.5$. The conductance is normalized by $G_{0}=Wp_F/\pi $ 
where $p_F$ is the Fermi wavevector in the incident side.

\subsubsection{Semimetalic phase ($\lambda_R > \Delta$)}

When $\lambda_R = 1.7$ (Fig. \ref{gc}(a), red curve), the system is in the semimetallic phase.
The charge neutrality (particle-hole symmetric) point on the transmitted
side is located at $V_0/\Delta=E/\Delta+1=1.5$.
The conductance \textit{vanishes} at this point. Above this
value of $V_0$, the conductance shows a singularity at $V_0/\Delta = E/\Delta-1
+2 \lambda_R/\Delta = 2.9$ (discontinuity in the first derivative). Here, the Fermi
energy touches the lowest energy band: $|--\rangle$ on the transmitted side.
Above this value, the number of
energy bands contributing to the conductance is doubled. The conductance
shows an abrupt increase of slope (a kink) at this point (see also Appendix).

Below the neutrality point ($V_0/\Delta=1.5$), the conductance shows a peak at $V_0=0$, then turns to a slow and monotonic decrease
as $V_0$ is decreased. This feature does not seem to resemble its behavior
above the neutrality point. Decreasing $V_0$ from $V_0=0$, the Fermi energy
touches the highest energy band: $|++\rangle$ at 
$V_0/\Delta= E/\Delta-1-2\lambda_R/\Delta = -3.9$. But the conductance curve does not show any singularity here.

This asymmetric behavior is a fingerprint of the unique band structure of
Kane-Mele model. At $V_0/\Delta=2.9$, the $|--\rangle$ band touching the Fermi energy
has the same band index $\beta=-1$, consequently the same symmetry as the
incident state: $|+-\rangle$. Therefore, as soon as this state becomes
available for transport, transmission occurs, typically, \textit{at the
normal incidence}, leading to an abrupt increase of the conductance. 
On the other hand, 
at $V_0/\Delta=-3.9$, the $|++\rangle$ band touching the Fermi surface
has the symmetry opposite to that of the incident state. 
Therefore, no
transmission occurs at the normal incidence via the $|++\rangle$ band, and
the conductance curve bears only a gradual change.
(see also Appendix A).

Note also that 
perfect reflection in the semimetallic phase is limited to
normal incidence, and
the conductance takes generally a finite value due to
contribution from non-zero incident angle, $\phi\ne 0$,
to the \textit{r.h.s.} of Eq. (\ref{LF}).
In the case of bilayer graphene, the conductance
curve shows similar features, with two shoulders only on the $V_0>0$ side.

\subsubsection{Topological gap phase ($\lambda_R<\Delta$)}

When $\lambda_R = 0.6$ (Fig. \ref{gc}(c), blue curve), the system bears a band gap. 
Naturally,
the conductance vanishes identically in the gap region: 
$-\Delta < E - V_0 <\Delta-2\lambda_R$, 
i.e., $0.7<V_0/\Delta<1.5$. 
The $|--\rangle$ band is always \textit{activated} for transport in the $pn$-regime:
$V_0/\Delta>E/\Delta+1=1.5$,
showing no singular behavior in the conductance curve.

\subsubsection{At the phase boundary ($\lambda_R = \Delta$)}

At the phase boundary (Fig. \ref{gc}(b), green curve), 
two bands with $\beta = -$ are combined, and a pair of
linearly dispersing energy bands, i.e., a Dirac cone appears.
The conductance curve is similar to that of monolayer graphene without
spin-orbit interaction. \cite{cayssol09}
The conductance vanishes at the charge-neutrality point $V_0=1.5$,
and shows again no singularity on the $pn$-side

\subsection{Fano factor}

Fano factor $F$ associated with the potential step is obtained 
\cite{cayssol09} from the transmission probability $T$ as
\begin{align}
F = \left. \int_{-p_F}^{p_F} dp_y T(1-T)
\right/
\int_{-p_F}^{p_F} dp_y T.
\end{align}
The lower panel of Fig. \ref{gc} shows the Fano factor as a function of $V_0$
for different values of $\lambda_R$.
One of the specific features is that
the Fano factor shows, independently of the value of $\lambda_R$,
a peak structure at $V_0=E+\Delta = 1.5$,
with a maximal value of $F$ ($=1$).
This is, of course, partly related to the fact that 
the conductance vanishes at this point.
However, recall also that 
in monolayer graphene without SO interaction,
the Fano factor remains structureless
at the corresponding point $V_0=E$ \cite{cayssol09},
with a value of $F \approx 0.1$.
Such characteristic suppression of shot noise is spoiled by the SO effects,
even when $\lambda_R = \Delta$, where Dirac cone reappears.
A zero of the conductance leads, trivially, to a peak in the Fano factor.
Very contrastingly, the charge conductance is not qualitatively affected by the SO effects.

The Rashba dominant regime exhibits some
anomalous features: 
a cusp in the conductance at $V_0/\Delta = 2.9$ 
for $\lambda_R/\Delta = 1.7$ (case (d) in Fig. \ref{gc}) 
and
an enhancement of the Fano factor.
in the regime of perfect reflection. 
This corresponds to $1.5 < V_0/\Delta < 2.9$ for $\lambda_R/\Delta = 1.7$.
In the region of perfect reflection,
evanescent modes are dominant in transport at a finite incident angle.
The Fano factor is enhanced by such evanescent modes.
The Fano factor increases with the increase of Rashba SO coupling 
in this regime of perfect reflection.
In the balanced case,
$\lambda_R =\Delta$ (Fig. \ref{gc}(e)), 
the conductance curves do not differ significantly from the case of no SO
effects ($\lambda_R=\Delta=0$).
In contrast, when intrinsic SO coupling dominates the Rashba term, 
i.e., in Fig \ref{gc}(f),
the conductance vanishes within a finite range of $V_0$, corresponding to
the gap.
The Fano factor is not well-defined, therefore, not plotted in this regime of $V_0$. 

In the large and small enough potential region $|V_0/\Delta| \gg 1$ and for arbitrary $\lambda_R$,
the Fano factor takes a value $F \approx 0.1$, 
which is roughly the same as the Fano factor without SO effects, since the transmitted state is free from evanescent modes in this region.

\begin{figure}
\includegraphics{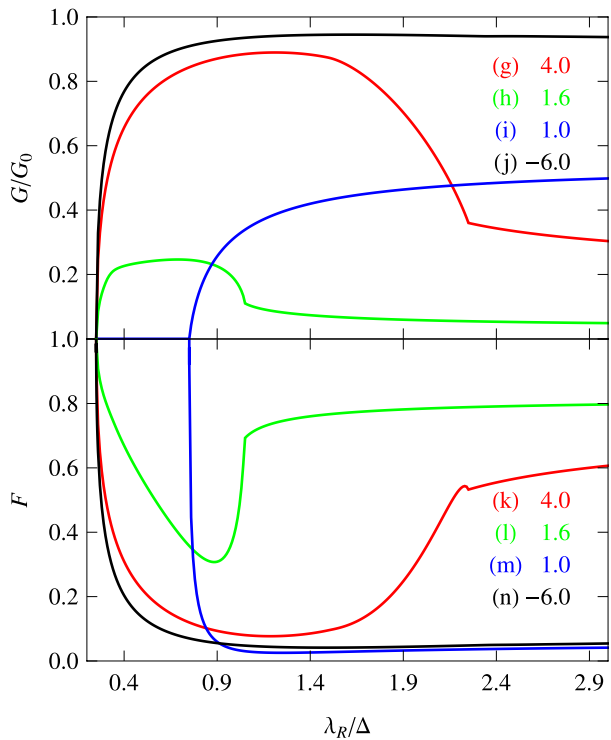}
\caption{Charge conductance (upper)  vs. Fano factor (lower) 
as a function of Rashba SO coupling. 
Incident energy is set to be $E/\Delta=0.5$.
Each curve corresponds to different values of $V_0/\Delta$.
}
\label{chcd_rshb}
\end{figure}

\subsection{Crossover effects in conductance and Fano factor}

The charge conductance and the Fano factor are shown in Fig. \ref{chcd_rshb} 
as a function of Rashba SO coupling $\lambda_R/\Delta$.
At normal incidence, we have seen a crossover 
from perfect reflection to perfect transmission 
in the $pn$-regime: $V_0/\Delta > 1.5$
(curves (i) and (ii) in the lower panel of Fig. \ref{rpcolor}). 
As for transport properties,
perfect transmission at normal incidence is replaced by
a broad maximum near $\lambda_R = \Delta$
(see cases (g) and (h) in the upper panel of Fig. \ref{chcd_rshb}).
The maximum always appears in the $pn$-regime.
However, unlike the normal incident case,
the maximum of charge conductance is not unity 
and is not precisely located at $\lambda_R = \Delta$ 
as a consequence of angular integration.

The charge conductance becomes smaller with the decrease of the $V_0$
on the $pn$-side (e.g., cases (g) and (h) in the upper panel of Fig. \ref{chcd_rshb}),
and actually vanishes at $V_0/\Delta=1.5$ (not plotted).
This is because the number of propagating states on the transmitted side
reduces with the decrease of the $V_0$
(and vanish at $V_0/\Delta=1.5$). 
A cusp of conductance appears also appears in this regime
at the band edge of $|--\rangle$.
When $\lambda_R/\Delta$ is larger than this value,
the Fermi energy intersects with only $\beta=+$ band.
As a result, perfect reflection occurs at the normal incidence,
leading to smaller values of conductance.

On the other side, i.e., in the $nn$-regime: $V_0/\Delta < 1.5$ (e.g., cases (i) and (j)),
maximum does not appear,  
because perfect reflection does not occur at the normal incidence 
(see (iii) and (iv) in the lower panel of Fig. \ref{rpcolor}).
When $V_0/\Delta = 1$ (case (i)) ,
charge conductance vanishes in 
$\lambda_R/\Delta \leq (1 - E/\Delta + V_0/\Delta)/2 = 0.75$, 
where the system is gapped on the transmitted side.

Dependence of Fano factor on Rashba SO coupling is
shown in the lower part of Fig. \ref{chcd_rshb}.
Interestingly, the lower panel looks almost the upside down image of
the upper panel.

\section{Spin transport}
\label{spintransport}

Here we investigate the spin transport generated by an electrostatic
potential step in the presence of spin-orbit effects. The potential step splits
the graphene sample in two pieces characterized by distinct carrier
densities. The spin is drifted along the interface and the spin current is
therefore transverse to the applied electric field. This spin Hall effect
(SHE) is a mesoscopic analog of the bulk spin Hall effect occurring in
homogeneous electron or hole doped semiconductors \cite{Murakami03,Sinova04}. 
The present effect requires both spin-orbit coupling and a step in the
carrier density. 
The spin Hall current localized in the vicinity of interface appears also in 2DEG with Rashba spin-orbit interaction
\cite{adagideli05}.

\begin{figure}
\includegraphics{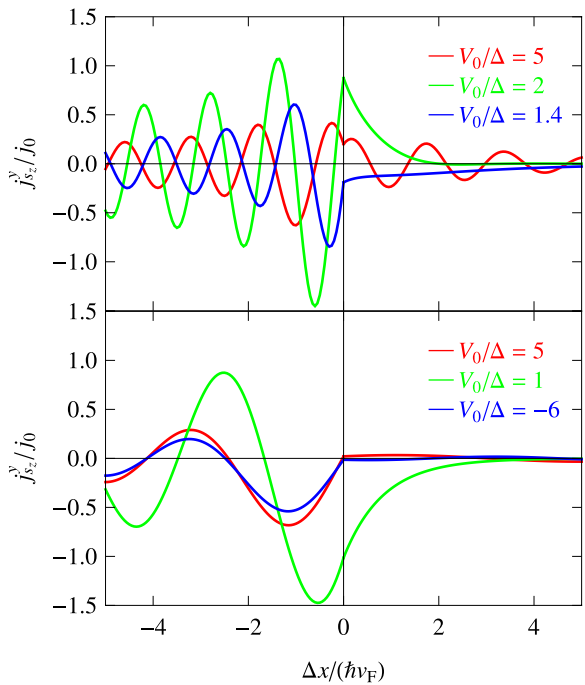}
\caption{Spatial distribution of spin current density. 
$\lambda_{R}/\Delta =2 \ (0.5)$ 
in the upper (lower) panel.
Spin Hall current is localized in
the vicinity of the junction. 
$j_0 = \Delta/(2\pi L)$.}
\label{jsR}
\end{figure}

\subsection{Local spin current density}

The spin flows along the $y$-axis, namely along the interface $x=0$ defined
by the potential step. The total spin current density, 
\begin{align}
j_{s_z}^{y}(\bm{x})
	&=\int_{-p_{F}}^{p_{F}}\frac{dp_{y}}{2\pi }
	WJ_{s_z}^{y}(%
\bm{x})
 \nonumber \\
 &=
	\displaystyle{\frac{\hbar W}{2}}
	\int_{-p_{F}}^{p_{F}}\frac{dp_{y}}{2\pi }
	\Psi^\dagger(\bm x) \sigma _{y}s_{z}
	\Psi(\bm x),
\label{jszlocal}
\end{align}
results from the summation of the single-channel currents $J_{s_z}^{y}(\bm{x%
})$ over all possible transverse modes labeled by their momenta $p_{y}$.
This local current is maximal near the interface and decays when the
distance $x$ is increased 
(Fig. \ref{jsR}). 
Note that the spacial dependence of spin current density
$j_{s_z}^{y}(\bm{x})$ is determined by a subtle interplay
between the direct and cross terms. When both transmitted waves are propagative, the (singlechannel) crossed
spin current oscillates as a function of the distance $x$ from the
interface. Of course, these oscillations are damped when integrated over all
possible transverse momenta $p_{y}$ (see Fig. \ref{jsR}) but this decay is a lonf ranged power law rather 
than an exponential decay.

When one or two transmitted wave(s) is/are evanescent, even the
singlechannel current decays exponentially when the distance $x$ from the
interface is increased. As a result, the total spin current density decays
very abruptly with $x$ (see Fig. \ref{jsR}).

\subsection{Spin conservation}

The same situation occurs for the $x$-component of spin current which comes only from the cross 
term, i.e. from Eq. (\ref{cross_s}). Moreover this contribution, due to spin torque in the presence of Rashba SO coupling,
vanishes after integration over incident angle. The spin transport
occurs, therefore, only in the direction parallel to the interface. Besides,
the continuity of spin current-density is recovered after this angular
averaging: 
\begin{equation*}
\int_{-p_{F}}^{p_{F}}\frac{dp_{y}}{2\pi }\left( {\frac{\partial \left \langle \rho _{s_{z}}%
 \right \rangle}{\partial t}}+\mathrm{div} \left \langle \bm{J}_{s_{z}}  \right \rangle \right) =0.
\end{equation*}%
The spin current density, defined as Eq. (\ref{js_def}), is thus considered
to be a conserved quantity in the sense of Eq. (\ref{spinconservation}), and free
from the usual issue of defining a spin current.

\subsection{Results}

\begin{figure}
\includegraphics{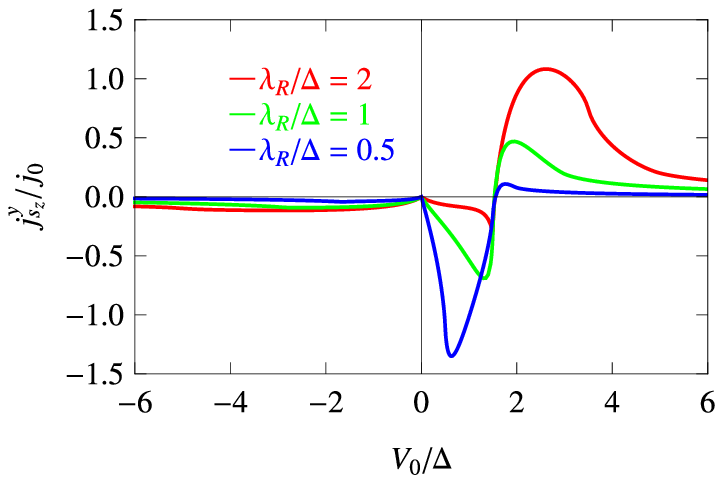}
\\[1em]
\includegraphics{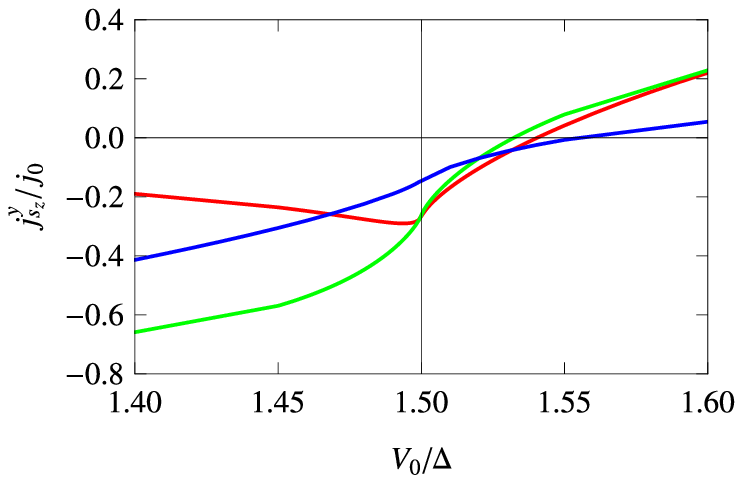}
\caption{Spin transport in the Kane-Mele p-n junction for different values
of $\protect\lambda _{R}$. The p-n junction induces $s_{z}$ spin current in
the direction \textit{parallel} to the interface. Such a spin Hall current
is plotted at $x=0$ as a function of $V_0$.}
\label{js_V}
\end{figure}

Fig. \ref{jsR} shows the spatial distribution of spin
current density parallel to the interface, 
i.e., $y$-component of spin current is plotted as a function of coordinate $x$ normal to the interface for different values of Rashba SO coupling: 
$\lambda_R/\Delta=2$ (dominant Rashba phase) for the upper panel of Fig. \ref{jsR}, 
and $\lambda_R/\Delta=0.5$ (topological gap phase) for the lower panel Fig. \ref{jsR}. 
 
In the upper panel, $1.5<V_0/\Delta<3.5$ corresponds to the
regime of perfect reflection, 
and the absolute value of spin current is large compared with other cases. 
The spin current is localized in the vicinity of interface, 
explicitly manifesting that spin is carried by evanescent modes
 (localized in the $x$-direction but propagating in the $y$-direction). 
It also shows a damped oscillatory behavior for a larger value of $V_0$ such as $V_0/\Delta=5$ in the upper panel. 
Such damped oscillation is an incarnation of cross terms between evanescent and propagating modes. 
The lower panel of Fig. \ref{jsR} corresponds to the topological gap phase, 
in which the spin current takes a large negative value when Fermi energy is in the gap on the transmitted side.

Fig. \ref{js_V} shows spin Hall current 
at $x=0$.
Let us compare it with the charge conductance shown in Fig. \ref{gc}, both
represented as a function of $V_0$. 
The two curves show indeed quite contrasting behaviors. At $V_0=0$ (in the
absence of a junction) the charge conductance show a maximum (peak). In
contrast, the spin current vanishes at $V_0=0$, reflecting the fact that here,
spin transport is a mesoscopic effect due to the presence of interface.

Both magnitude and sign of the spin current is tuned by $V_0$. The direction of spin current is opposite
between inter-band tunneling ($V_0>E+\Delta$) and 
intra-band tunneling ($V_0<E+\Delta$) cases
(compare the two panels of Fig. \ref{js_V}).
\footnote{Some anomalous behaviors can be seen
in the vicinity of $V_0=E+\Delta$, though.}
The latter includes also the case of metal-insulator junction.
We can confirm this explicitly from Eqs. (\ref{jszy_direct}) and (\ref{cross_s}) with $E \to \tilde E = E-V_0$.
The direct term of spin Hall current is proportional to $E-V_0+\Delta$, therefore the direction changes near $V_0=E+\Delta$.
The sign of the crossed term is opposite from the direct terms and enhanced in the vicinity of Dirac point $V_0=E+\Delta$, resulting finite spin current at $V_0=E+\Delta$.

In the dominant Rashba phase ($\lambda_R>\Delta$), two quadratic bands touch
at the neutrality point $E=-\Delta$. On the transmitted side, this
corresponds to $V_0 = E+\Delta \equiv V_n$. 
At this value of $V_0$, the charge
conductance vanishes. The spin conductance changes its sign in the vicinity
of neutrality point, but remains finite precisely on that point 
(see the lower panel of Fig. \ref{js_V}). 
This is
again due to the evanescent modes. 
The spin current takes a large
positive value above the neutrality point: $V_0>V_n$. 
In the topological gap
phase ($\lambda_R<\Delta$), the charge conductances vanishes in the gap: $%
E-\Delta<V_0<E+\Delta=V_n$, whereas the spin current is enhanced in the
gap, taking a large negative value.

\begin{figure}
\includegraphics{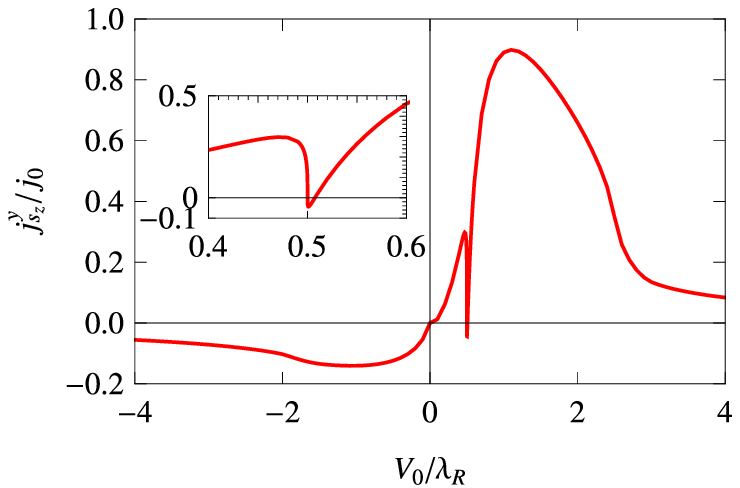}
\\[1em]
\includegraphics{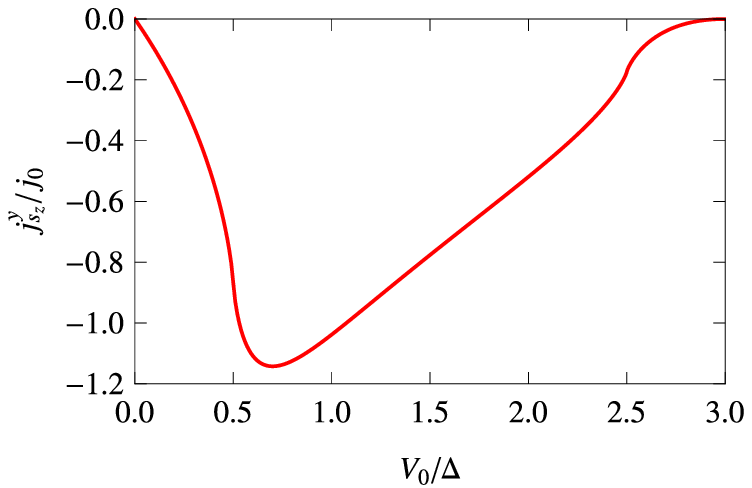}
\caption{Mesoscopic spin Hall effect in the absence of topological mass term: $\Delta=0$ (Rashba SO coupling: $\lambda_R=0$) in the upper (lower) panel. 
Spin Hall currents at $x=0$ are plotted as a function of $V_0$ for $E/\lambda_R=0.5$ with $\Delta=0$, $j_0=\lambda_R/(2\pi L)$ and $E/\Delta = 1.5$ with $\lambda_R=0$, $j_0=\Delta/(2\pi L)$.}
\label{gsR}
\end{figure}
The upper panel of Fig. \ref{gsR} 
reveals two different natures of mesoscopic
spin Hall effect by studying, separately, the $\Delta=0$ and $\lambda_R=0$
cases. 
Fig. \ref{gsR} shows that the mesoscopic spin Hall current flows
actually in the absence of topological mass term: $\Delta=0$. The spin Hall
current is enhanced in the regime of perfect reflection: $E<V_0<E+2\lambda_R$ 
($0.5 < V_0/\lambda_R < 2.5$ in Fig. \ref{gsR}).
This corresponds to the regime of $V_0$ above the neutrality point, where the
spin conductance takes a large positive value. 
The lower panel of Fig. \ref{gsR} shows, on
contrary, $J_{s_z}^y$ as a function of $V_0$ in the absence of Rashba SOC: $\lambda_R=0$. 
Clearly, the spin current is enhanced, when Fermi
energy is in the gap on the transmitted side: $E-\Delta<V_0<E+\Delta=V_n$ ($0.5 < V_0/\Delta < 2.5$ in Fig. \ref{gsR}),
i.e., in the situation of metal-insulator junction.
In the region of $V_0<0$ and $V_0>2E$,
the spin current vanishes because only propagating mode appears.
The spin degeneracy remains in the absence of Rashba SOC,
as a result, the crossed term also vanishes.

On the transmitted side, the enhancement of spin Hall current thus occurs for two reasons: (i) perfect reflection in the dominant Rashba phase
(on the $V_0>V_n$ side), (ii) due to the topological gap (on the $V_0<V_n$
side). In the two cases, evanescent modes play a dominant role in the
solution of scattering problem at the junction. The enhancement of spin Hall
current along the interface thus has two different natures, both related to
evanescent modes. Depending on which side of the neutrality point Fermi
energy on the transmitted side is, the enhanced spin current flows in the
opposite directions.

\begin{figure*}
\begin{minipage}{0.49\hsize}
\includegraphics{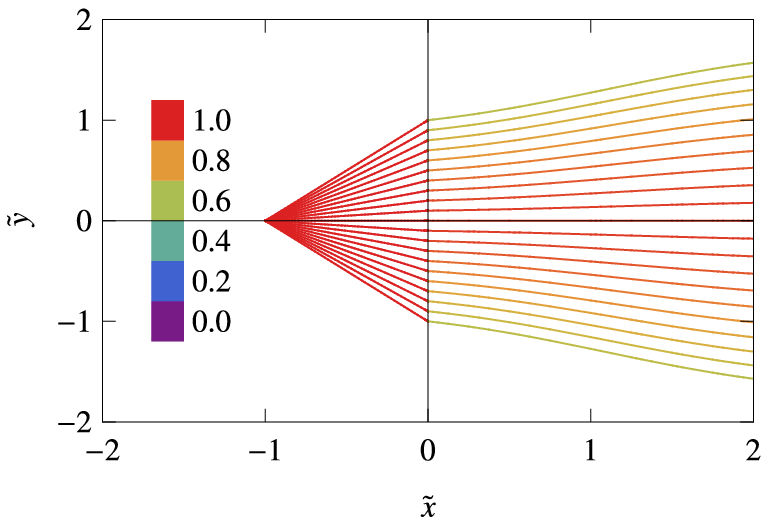}
\end{minipage}
\begin{minipage}{0.49\hsize}
\includegraphics{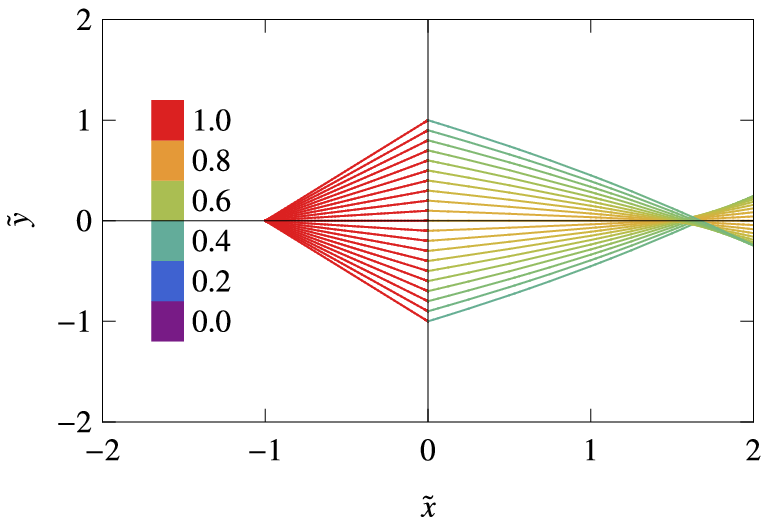}
\end{minipage}
\caption{Refraction of the electron beams by the potential step in the topological gap phase ($\protect\lambda _{R}/\Delta =0.5$%
). Left panel (a): intra-band
tunneling ($V_0/\Delta =-2$). Right panel (b): inter-band tunneling ($V_0/\Delta=2$). Spatial coordinates: $\tilde{x}=x\Delta /(\hbar v_{F})$, $\tilde{y}%
=y\Delta /(\hbar v_{F})$. Fermi energy: $E/\Delta =0.5$.}
\label{lens_topo}
\end{figure*}
\begin{figure*}
\begin{minipage}{0.49 \hsize}
\includegraphics{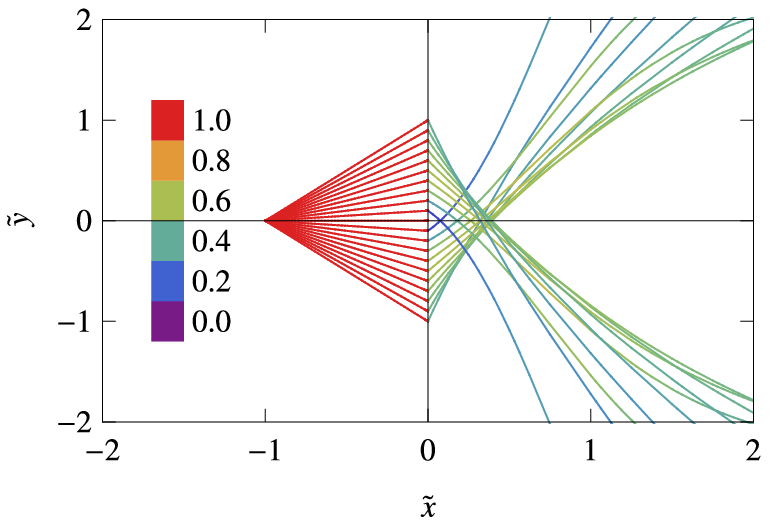}
\end{minipage}
\begin{minipage}{0.49\hsize}
\includegraphics{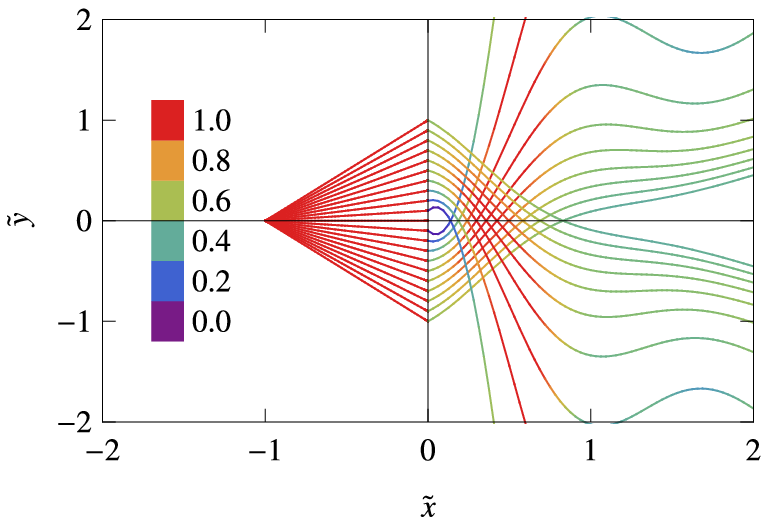}
\end{minipage}
\caption{Electron Veselago lens --- semimetallic phase ($\protect\lambda%
_R/\Delta = 2$). 
Spatial coordinates: $\tilde x = x\Delta/(\hbar v_F)$, $%
\tilde y = y\Delta/(\hbar v_F)$. 
Fermi energy: $E/\Delta =
0.5$. 
The height of potential step $V_0$ is chosen such that (a) left panel: $%
V_0/\Delta = 2$, (b) right panel: $V_0/\Delta = 3.5$, both corresponding to the
inter-band tunneling case. Case (b) corresponds to the opening of the lowest
energy channel.}
\label{lens_bi}
\end{figure*}

\section{Electron Veselago lensing}

The $p-n$ junction in graphene is expected to serve as an electronic version
of "Veselago lens" \cite{Vadim}. 
The charge current distribution: ${\bm j}_{\rm c}
({\bm x})=(j_{\rm c}^x ({\bm x}), j_{\rm c}^y ({\bm x}))$, can be used for imaging the electronic flow around the potential step \cite{cserti2007}. 

Let us imagine an electronic wave packet emitted from a point ${\bm x}%
=(-a,0)$ with $a>0$. If the wave
packet has a center-of-mass momentum ${\bm p}=(p_x, p_y)$ with $p_y/p_x=\tan
\phi$, then the wave packet will be incident at the $p-n$ junction (located at 
$x=0$) at $y=a\tan \phi$. Let us consider the trajectory of this wave packet
after it goes through the $pn$-junction.

Here, instead of following explicitly the dynamics of such a wave packet, we
calculate directly the stationary charge current distribution on the transmitted side,
using Eq. (\ref{d-c}). 
Then, we consider stream lines of the vector field $\bm j_{\rm c} (\bm x)$.
Once ${\bm j}_{\rm c} ({\bm x})$ is known, the following differential
equation: 
\begin{align}
\frac{dy}{dx} = \frac{j_{\rm c}^{y} (\bm x)}{j_{\rm c}^{x} (%
\boldsymbol{x})},  \label{deq}
\end{align}
determines the locus of a stream line under a given boundary condition, say, 
$y(x=0)=a\tan\phi$. 
Equation (\ref{deq}) determines, in turn, the trajectory of
the wave packet emitted from ${\bm x}=(-a,0)$. 
As we have seen in
Eqs. (\ref{jc_bulk}), (\ref{js_bulk}), (\ref{jc_bulkeva}), (\ref{jc_bulkevay}), (\ref{cross_c}),
$j^{y}_{\rm c}$ has no $y$-dependence, whereas $j^{x}_{\rm c}$ has no spatial dependence due to charge conservation. Therefore, with a given boundary
condition, Eq. (\ref{deq}) can be trivially integrated to give, 
\begin{align}
y(x) = a \tan\phi + \frac{1}{j^{x}_{\rm c}} \int_0^x dx^{\prime}{j^y_{\rm c}(x^{\prime})}.
\label{traj}
\end{align}
Repeating the same procedure for different values of incident angle $\phi$,
one can draw a set of stream lines visualizing the vector field ${\bm j}_{\rm c} ({\bm x})$. Focusing of such stream lines can be regarded as an
electronic version of optical lens. 
Figs. \ref{lens_topo} and \ref{lens_bi}
demonstrate such electron lens realized at the Kane-Mele $p-n$ junction.
A color code specifies the strength of current flow along each trajectory.

\subsection{Topological gap phase}

In the topogical gap phase, the refraction properties are quite similar to the ones studied previously for graphene $p-n$ junctions in the absence of spin-orbit interaction \cite{Vadim}. Indeed the $y$-component of the current density changes its sign on crossing the $p-n$ junction, thereby realizing the negative or Veselago-like electronic refraction. In presence of SOC, the system does not longer show perfect focusing (Fig. \ref{lens_topo}.b).  

In contrast for a $n-n$ junction (intraband transmission) the $y$-component of the current has the \textit{same} sign on both sides of the junction, indicating that the refractive index is \textit{positive}. As a result, the outgoing electron beam is divergent (Fig. \ref{lens_topo}.a).

\subsection{Semimetallic phase}

More interesting are the refraction properties of the semimetallic phase. Indeed the evanescent modes manifest themselves by the bending of 
the electronic rays on the transmitted side (Fig. \ref{lens_bi}). Moreover no transmission is allowed at the normal incidence which yields a shade area behind the origin O$(0,0)$ (Fig. \ref{lens_bi}). 

Figure \ref{lens_bi} shows Veselago-like electron lens in the semimetallic
phase ($\lambda_R/\Delta = 2$) for two distinct steps $V_0$: (a) $V_0/\Delta = 2$ (left panel), (b) $%
V_0/\Delta = 3.5$ (right panel), both corresponding to a $p-n$ junction (inter-band tunneling).

In case (b), the Fermi energy is touching the top of $|--\rangle$-band, which happens when $V_0=E -
\Delta + 2\lambda_R$. As the Fermi energy approaches the the top of $%
|--\rangle$-band, the \textit{cross} term between propagating (due to $%
|-+\rangle$-band) and evanescent (due to $|--\rangle$-band) modes plays a
significant role. The meandering stream lines in case (b) is a consequence
of such "cross term transport". It should be also noticed that the shade area is much
pronounced in case (b). Indeed, one can observe that the refractive index turns to be
positive in the vicinity of the shade area. Due to perfect reflection at the
normal incidence, the $x$-component of current density is virtually zero for
a small incident angle $\phi$. As a result, the electron beam is strongly
refracted, with divergent $j_y/j_x$ on the transmitted side (as $%
\phi\rightarrow 0$). This leads to the formation of shade area (see Fig. \ref%
{enlarged} for detailed plots).

\begin{figure}
\begin{center}
\includegraphics{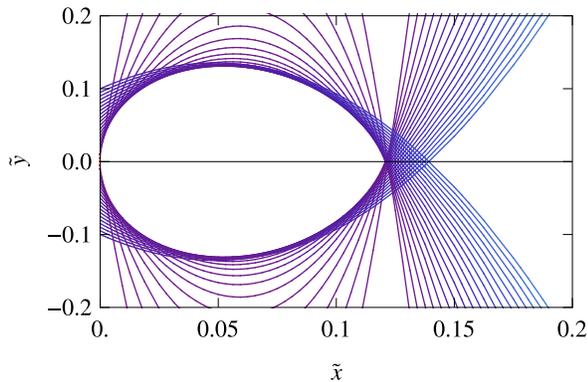}
\end{center}
\caption{Detailed plots of the shade region of Fig. \protect\ref{lens_bi}
(b): $\protect\lambda_R/\Delta = 2$, $V_0/\Delta =3.5$. Other parameters are
also the same.}
\label{enlarged}
\end{figure}

Study of electron lens behavior thus reveals rich mesoscopic transport
properties of the Kane-Mele $pn$-junction. It should be underlined that
this unique mesoscopic transport is carried out by the evanescent modes and the
cross terms. The former enables transport along the edge even when the Fermi
level is in the gap on the transmitted side \cite{Nori}. 

Finally the strinking difference between the behaviors of the topological gap phase (Fig. \ref{lens_topo}) and the semimetallic phase (\ref{lens_bi}) might be observed by 
scanning probe measurements similar to those successfully implemented on top of ballistic two-dimensional electron gases \cite{topinka2000,topinka2001}.

\section{Conclusion}
We have studied theoretically charge and spin transport at a potential step (both $n-n$ and $p-n$ junctions) within the Kane-Mele of graphene.
We have highlighted the role of reflection symmetry associated with the band index $\beta$ in the crossover 
from {\it perfect} reflection to transmission while tuning the Rashba coupling $\lambda_R$. We have also computed experimentally measurable quantities 
such as conductance and Fano factor. 

Due to the multiband character of the model, one incident electron yields two distinct transmitted quasiparticles. The spin Hall current, which is mainly localized in the vicinity of interface, results from the superposition of two types of contributions: (i) direct terms involving one kind of transmitted quasiparticles and (ii) cross terms describing interferences between the two kinds of transmitted quasiparticles (Sec V). The direct terms were shown to carry no
net $s_{z}$-spin current when associated with a propagative wave whereas evanescent waves carry a finite spin current. In contrast the crossed terms always contribute to
spin transport regardless of the nature of the waves.The interplay between those direct and cross terms is also important for charge transport (Sec. IV).

Moreover the electronic flow exhibits a large variety of patterns (Sec. VI). In particular a dominant Rashba SOC (semimetallic phase) leads to curved rays owing to the presence of evanescent states whereas rays are still straight for dominant intrinsic SOC (topological gap phase). Note that in a monolayer graphene without SOC, stream lines are straight and refracted only at the interface.  In principle, it should be possible to identify those contrasted shapes by scanning a charged tip above the graphene flake as it was performed for two-dimensional electron gases in GaAs heterostructures \cite{topinka2000,topinka2001}. Finally detecting a clear fingerprint of the role of SOC in transport measurements in graphene seems not impossible but  difficult, since the magnitude of SOC is small in graphene, at most, on the order of $\sim 1 \rm K$. \cite{gmitra09}. An alternative way to probe such unique transport characteristics of $p-n$ junction may be to use materials with stronger SOC, such as HgTe/CdTe heterostructures.

\acknowledgments
K.I. and A.Y. are supported by KAKENHI (K.I.: Grant-in-Aid for Young Scientists under Grants No. B-19740189 and A.Y.: No. 08J56061 of MEXT, Japan). JC is supported by the 7$^{th}$ European Community Framework Programme under the contract TEMSSOC (joint program between UC Berkeley and the Max-Planck-Institut für Physik Komplexer Systeme in Dresden).

\appendix*

\begin{figure}
\includegraphics{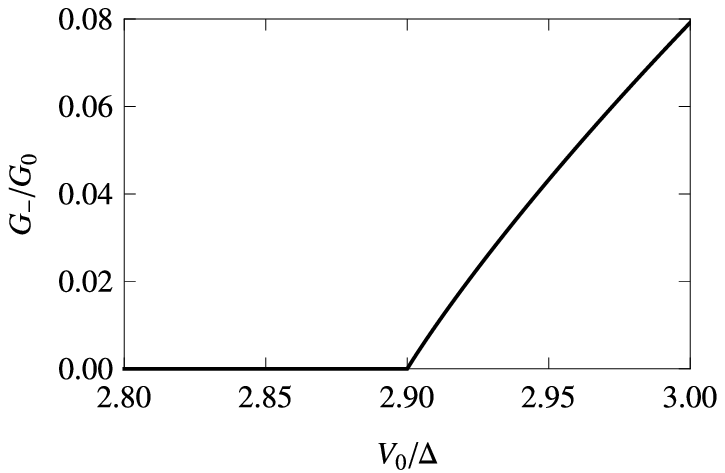}
\\
\hspace{-1em}
\includegraphics{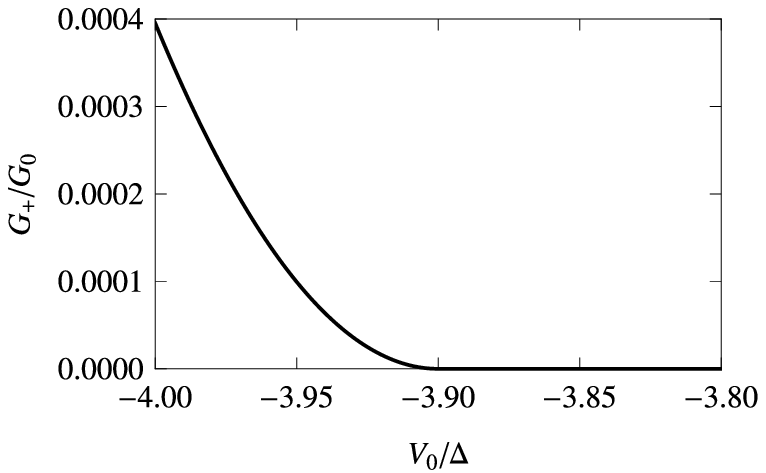}
\caption{Contribution of $|--\rangle$ band to charge conductance. The charge conductance $G_-$ ($G_+$) calculated from $j_-$ ($j_+$) is shown in the upper (lower) panel.
System's parameters are chosen such that $\lambda_R/\Delta=1.7,E/\Delta=0.5$.}
\label{gpm}
\end{figure}

\section{Asymptotic Behavior of Conductance}

The curve of $\lambda_R/\Delta=1.7$ in Fig. \ref{gc} illustrates the characteristic features of
charge conductance in the semimetallic phase: $\lambda_R>1$. The conductance
curve shows, say, at $\lambda_R = 1.7$, a kink structure at $V_0=E-\Delta+2\lambda_R=2.9$, 
on opening of the $|--\rangle$-channel
to transmission. The purpose of this appendix is to estimate the asymptotic
behavior of $G_{\rm c}$ in the vicinity of this singularity.

In Sec. \ref{charge_conductance} we estimated $G_{\rm c}$, by substituting $T(\phi)=1-|r(\phi)|^2$ with 
$r(\phi)$ determined by the continuity condition (\ref{cc}), into the
Landauer formula (\ref{LF}). Here, to reveal the nature of singularity at $%
V_0=E-\Delta+2\lambda_R$, we extract the contribution of $|--\rangle$%
-band to $G_{\rm c}$, and analyze its asymptotic behavior in the vicinity of the
singularity (see Fig. \ref{gpm}). 
Let us introduce $\delta V$, the height of potential barrier
measured from the singularity: 
\begin{align}
\delta V \equiv V_0-E +\Delta - 2\lambda_R.
\end{align}
When $\delta V \le 0$, the $|--\rangle$-band gives no contribution to the
charge current on the transmitted side, since the state is evanescent. When $%
\delta V > 0$, a propagating mode becomes possible with the momentum 
\begin{align}
\hbar v_F q_-  
= - \sqrt{2(\lambda_R-\Delta)\delta V - (\hbar v_F k \sin\phi)^2},
\end{align}
obtained from the energy conservation: $E_{\alpha\beta}(p_{\beta x},p_y)=E-V_0$, provided $|\phi| \le \phi_m$, where the critical angle $\phi_m$ is defined by, 
\begin{align}
(\hbar v_F k \sin \phi_m)^2 = 2(\lambda_R-\Delta) \delta V.
\end{align}
In the limit of $\delta V \rightarrow 0$, this reduces to, 
\begin{align}
\phi_m \sim \sqrt{2(\lambda_R - \Delta) \delta V}/(\hbar v_F k).
\end{align}
The charge current transmitted to the lowest energy band is given by 
\begin{align}
j_{-}(\epsilon_F,\phi) = -\frac{e v_F}{WL} \frac{2 |t_-|^2 \hbar v_F q_-}{%
E - V_0 + \lambda_R},
\end{align}
from Eq. (\ref{jc_bulk}).
Integrating over all incident angles, one finds 
\begin{align}
&\int_{-\phi_m}^{\phi_m} d\phi j_-(E,\phi) \sim 2 \phi_m
j_-(\epsilon_F,0)  \notag \\
&= \frac{8 \delta V}{E+\lambda_R} \sqrt{\frac{E+\Delta}{%
E-\Delta+2\lambda_R}}.
\end{align}
The charge current transmitted to the $|--\rangle$-band thus shows a \textit{%
linear} uprise (proportional to $\delta V$) when $\delta V>0$. As a result,
the charge conductance shows an abrupt increase of slope at $%
V_0=E+\Delta$. Note that $t_-(E,0) \ne 0$ thanks to the same
symmetry, i.e., the same $\beta$ ($=-1$), of the $|--\rangle$-band as that
of the incident energy band: $|+-\rangle$.

Similarly, we investigated the asymptotic behavior of $G_{\rm c}$ in the vicinity
of the opening of $|++\rangle$-channel at $V_0= E-\Delta-2\lambda_R$. However,
since the $|++\rangle$-band (final state) has the opposite symmetry
(opposite $\beta$) to the initial state: $|+-\rangle$. Due to this mismatch
of symmetry, transmission coefficient $t_+(\phi=0)$ vanishes even when $V_0 \le
E -\Delta - 2\lambda_R$. As a result, the leading order
contribution to the charge conductance starts at second order, i.e., $%
\propto \delta V^2$ (see Fig. \ref{gpm}, right panel).

\bibliography{100907_MSHE}

\clearpage

\end{document}